# Real time noninvasive cancer diagnostics

Z.V. Jaliashvili,[1] T.D. Medoidze,[1] Z.G. Melikishvili,[1,*] K.M. Mardaleishvili[2], and J.J. Ramsden[3]

[1]*Department of Coherent Optics and Electronics, Institute of Cybernetics of the Georgian Academy of Sciences, 5 Sandro Euli St, 0186 Tbilisi, Georgia*

[2]*National Cancer Center of Georgia, Lisi Lake, 0177 Tbilisi, Georgia*

[3] *Department of Advanced Materials, Cranfield University, MK43 0AL, England*

**Abstract**
Laser illumination of tissue results in a characteristic fluorescence emission spectrum whose features depend on the type of tissue, viz., healthy, adenoma or malignant. Hence measurement of the fluorescence is potentially a rapid and reliable diagnostic method. We have applied the technique to thyroid tissues and found that judicious normalization of the fluorescence spectrum reveals a close correlation between spectral shape and tissue type. Hence the pathological state of the tissue can be unambiguously identified from the laser-induced fluorescence spectrum. The method is rapid and portable and is therefore eminently suitable for use during operations.

Most living tissue fluoresces in one way or another. The challenge is to harness the phenomenon in such a way that it is useful to the surgeon. Here we examine the laser-induced fluorescence of thyroid tissues. It turns out that the fluorescence spectra can be captured in such a way that unambiguous differentiation between normal (healthy), adenoma and malignant tumours can be made.

Many forms of pre-cancerous changes are difficult to detect and diagnose using conventional techniques, which require histological examination of biopsies obtained from visible lesions, or random surveillance biopsies [1,2]. These techniques not only often yield ambiguous results, but they are also slow. What is needed is a method for diagnosis that can be used during an operation, to reliably identify malignant tumours, adenomas and healthy regions within a second.

*Laser-induced fluorescence spectroscopy* (LIFS). Biological tissue is a turbid optical medium, in which light from an incident laser beam first undergoes Fresnel reflexion from the tissue surface. The rest of the beam is transmitted into the tissue and its fate is dominated by absorption or multiple scattering or both. The relation between absorption and scattering depends upon the wavelength of the laser radiation and the optical properties of the tissue. In the ultraviolet range the thyroid tissue is a strong absorber. The primary fluorescence centres for ultraviolet radiation are native fluorophors such as tryptophan, collagen, elastin and NADH [3].

---

* Correspondence concerning this article should be addressed to Dr. Zaza Melikishvili, Department of Coherent Optics and Electronics, Institute of Cybernetics, 0186 Tbilisi, Georgia. E-mail: z_melikishvili@posta.ge.

Thyroid tissue samples were obtained from 10 different individuals as solid chunks a few mm thick. Each sample was divided into two parts, one half going for histo-morphological examination and the other for the fluorescence measurements. The latter were placed in a spectroscopic quartz tube and illuminated with 10 ns, 337 nm pulses from a nitrogen laser at a repetition rate of 100 Hz. The energy per pulse was 0.04 mJ. The beam was focused into a spot 100 μm in diameter at the front surface of the tissue. Fluorescent emission was collected from the front surface and passed through the entrance slit of a DFS-452 double diffraction spectrograph/monochromator (1200 grooves/mm). Empty tubes neither absorbed nor emitted, but showed only Fresnel reflexion. A UNIPAN-233 selective nanovoltmeter (working in repetitive pulse mode) was used to read the output from the spectrograph photomultiplier. Emission spectra were recorded from 350 to 600 nm. Measurements were made at several different spots on each sample.

Examination of many samples of different types revealed that the fluorescence emission spectra had the following three features: a rather constant peak centred at 465 nm, a peak of variable intensity centred at 395 nm, and a rising background starting at about 420 nm and increasing towards shorter wavelengths. Figure 1 shows the fluorescence emission spectra from three different tissue types, normal (healthy), adenoma and malignant. The spectra have been normalized to give the same peak emission intensity at 465 nm. This normalization is an essential step for exploiting the full power of the LIFS technique.

The key observation for making the fluorescent emission into a surgically useful diagnostic technique was that malignant tumours, adenomas and healthy tissue have distinctly different relations between the intensities of these three features. Malignant cancerous tissue is dominated by the rising background, and adenomas by the peak at 395 nm.

Although very roughly speaking adenoma gives a spectrum intermediate between the spectra of healthy and malignant tissue, the different spectral lineshapes make it possible to unambiguously differentiate mixed healthy and malignant tissue from adenoma. This is made clear in Figure 2, which shows two samples of intermediate character. Indeed, the averaged spectra of the extreme forms (healthy and malignant) can be fitted as a linear combination to the mixed spectra and hence used to quantitatively determine the fractions of normal and malignant tissue in the sample, or, in effect, the pathological degree.

LIFS is a novel optical method developed to probe the state of living cells in situ without the need for tissue removal. It is useful for real-time, noninvasive medical imaging starting from the very early stages of disease. LIFS makes it possible to distinguish between the emission of native fluorophors in normal and different types of abnormal human cells. The fluorescence spectrum is analysed to provide qualitative and quantitative information about the states of biomolecules and their variations across the tissue. As LIFS operates on the molecular level, it is a robust and powerful tool for detection of different pathologies in optically accessible organs, without being influenced by other properties of the tissue, such as the age of its donor.

The rapidity of the diagnosis (within a fraction of a second) makes the technique eminently suited for quickly mapping the boundary between malignant and normal tissue during an operation. Presently the spectroscopy is being carried out in the optics laboratory using a large, cumbersome and expensive setup [4]; however, every component could be miniaturized and our next goal is to construct a small, portable unit that can be used directly in the operating theatre. We are also working on the generalization of the technique to all forms of cancer.

## References

1. Gurjar RS, Backman R.S, Perelman LT, Georggakoudi I, Badizadegan K, Itzkan I, Dasari RR, Feld, MS. Imaging human epithelial properties with polarized light-scattering spectroscopy. Nature Medicine 2001; 7: 1245–48
2. Cotran RS, Robbins SL, Cumar V. Robbins Pathological Basis of Disease (Philadelphia: W.B. Saunders, 1994)
3. Song JM, Jagannathan R, Stokes DL, Kasili PM, Panjehpour M, Phan MN, Overholt BF, DeNovo RC, Pan X, Lee RJ, Vo-Dinh T. Development of a fluorescence detection system using optical parametric oscillator (OPO) laser excitation for in vivo diagnosis. Technology in Cancer Research & Treatment 2003; 2: 515–24
4. Medoidze TD, Melikishvili ZG. Spectroscopy and dynamics of transitions in uv-excited Tm3+:YLiF4 laser systems. In: Arkin WT (Ed.) Focus on Lasers and Electro-Optics Research (New York: Nova Science Publishers, 2004)

| Feature[a] | | Interpretation |
|---|---|---|
| $I_{395}$ | Rising background[b] | |
| $\ll I_{465}$ | absent | normal (healthy) |
| $\sim I_{465}$ | absent | adenoma |
| shoulder | strongly rising towards the UV | malignant cancer |
| pronounced shoulder | weakly rising towards the UV | mixed malignant and healthy |

Table 1: Diagnostic drill for thyroid cancer. $I_\lambda$ represents feature intensity at wavelength $\lambda$.
[a] After normalizing the 465 nm peak to unity. [b] Starting at about 420 nm and increasing towards shorter wavelengths.

**Acknowledgement**. The authors thank the Collegium Basilea (Institute of Advanced Study), Basel, Switzerland for a travel grant (to JJR). All other funding was from internal sources (i.e. the authors' respective institutions).

## FIGURE LEGENDS

**Figure 1**. Fluorescence emission spectra from normal (healthy) (solid line), adenoma (dashed line) and malignant tissue (dotted line), as determined by histo-morphological examination. Each spectrum is a average of 2 measurements from each of 10 different samples. Spectra are scaled to give the same peak emission intensity at 465 nm.

**Figure 2**. Fluorescence emission spectra from two samples of mixed character, to which linear combinations of the extreme forms, healthy and malignant (also shown, as black solid and black dotted lines respectively) were fitted. Upper solid gray curve: spectrum from 70% healthy and 30% cancerous; lower solid gray curve: spectrum from 55% healthy and 45% cancerous; the corresponding fittings are shown as dashed black lines.

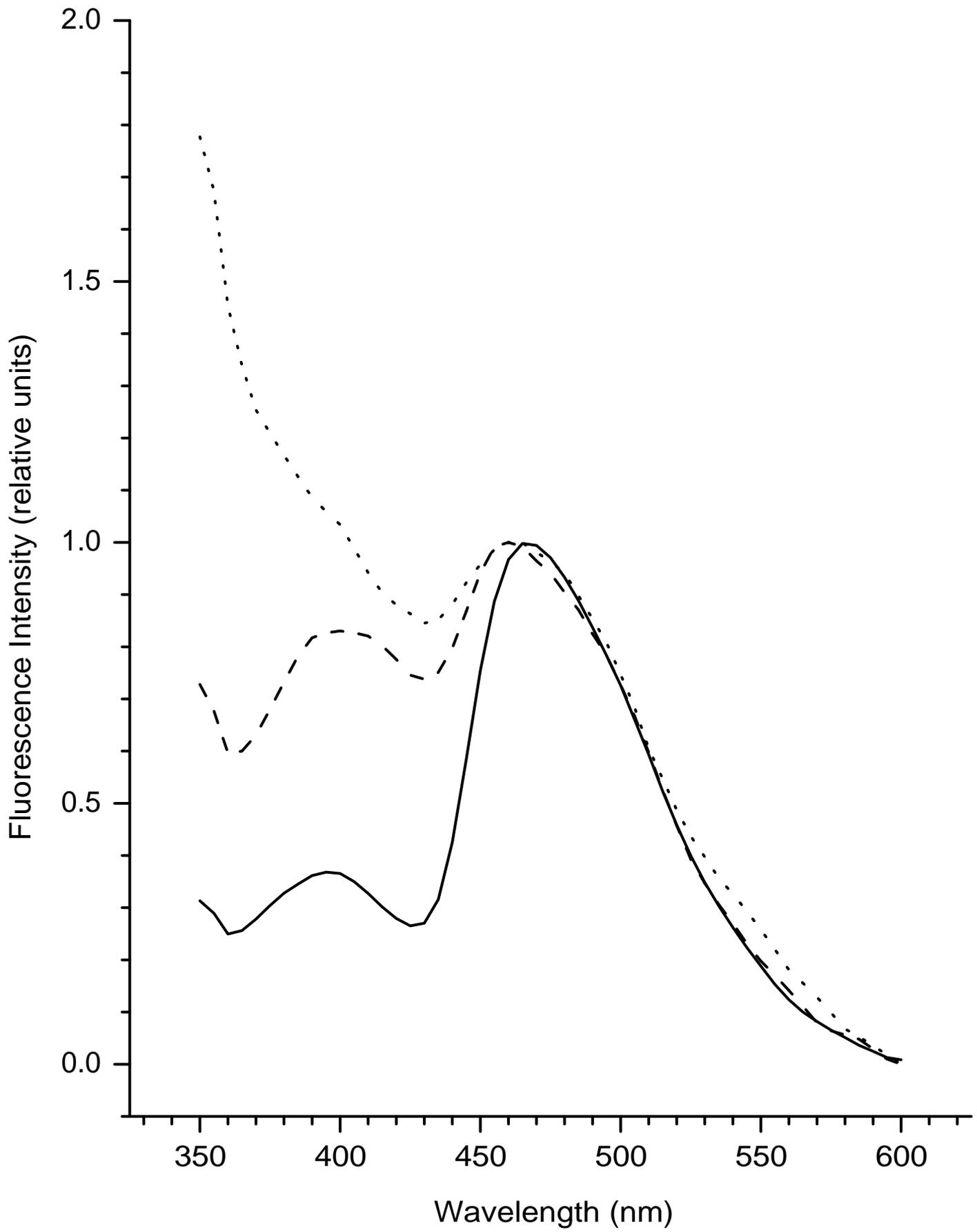

Fig. 1

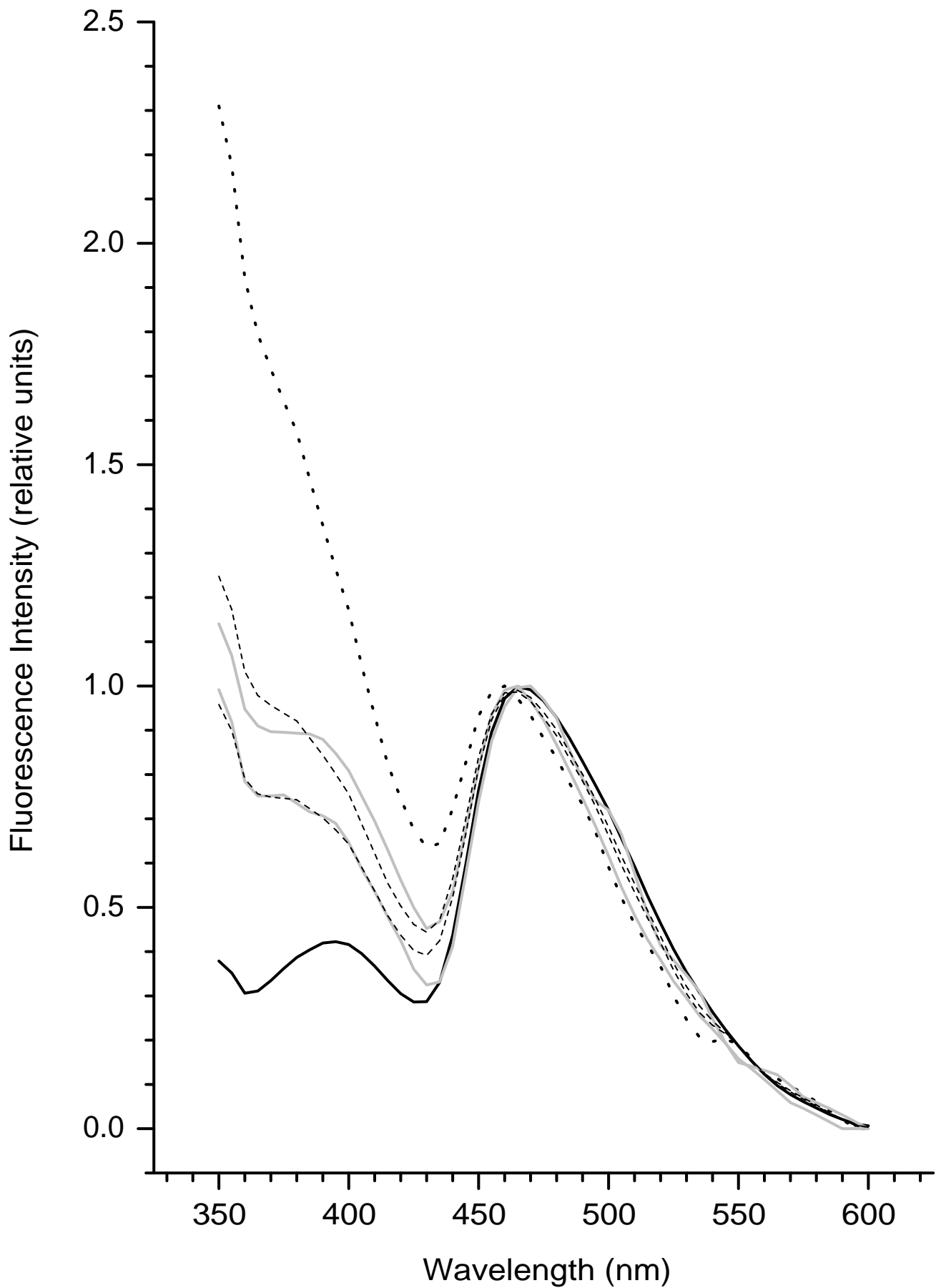

Fig. 2